\documentclass[conference]{IEEEtran}

\usepackage{algorithm,algpseudocode}
\usepackage{subfig}
\usepackage{graphicx}
\usepackage{amsmath,amssymb,amsfonts}
\usepackage{hyperref}
\usepackage{cleveref}
\usepackage{xcolor,pifont,xspace}
	
\usepackage{color, colortbl}
\usepackage{tikz}
\usepackage{amsmath}

\newcommand{\prohibitsign}{%
  \tikz[baseline=-0.5ex, scale=0.1]{
    \fill[red] (0,0) circle(1.8);
    \fill[white] (0,0) circle(1.2);
    \draw[line width=0.05cm, red] (-0.8,0.8) -- (0.8,-0.8);
  }%
}
\newcommand{\heartcheck}{%
  \tikz[baseline=-0.5ex, scale=0.12]{
    \fill[green!70!black] 
      (0,0) .. controls (0,1.5) and (-1.5,1.5) .. (-1.5,0.5)
              .. controls (-1.5,-0.5) and (0,-1.5) .. (0,-2)
              .. controls (0,-1.5) and (1.5,-0.5) .. (1.5,0.5)
              .. controls (1.5,1.5) and (0,1.5) .. (0,0);
    \draw[white, line width=1.2pt] (-0.7,-0.3) -- (-0.2,-0.9) -- (0.7,0.3);
  }%
}

\newcommand{\nadd}[1]{{\color{black}{#1}}}

\newcommand{\ournameNoSpace}{GoldenFuzz} 
\newcommand{\ourname}{\ournameNoSpace\xspace}

\newcommand{\newbugcount}{{five~}}

\usepackage{listings}
\usepackage{xcolor}

\lstdefinestyle{asmstyle}{
    language=[x86masm]Assembler,   
    basicstyle=\ttfamily\small,   
    keywordstyle=\color{blue},    
    commentstyle=\color{green!50!black}, 
    stringstyle=\color{red},      
    numbers=left,                 
    numberstyle=\tiny\color{gray},
    stepnumber=1,                 
    numbersep=10pt,               
    backgroundcolor=\color{gray!10}, 
    frame=single,                 
    tabsize=4,                    
    captionpos=b,                 
    breaklines=true,              
    breakatwhitespace=true,       
    showspaces=false,             
    showtabs=false,               
    showstringspaces=false,       
}

\lstdefinestyle{fancyasm}{
    language=[x86masm]Assembler,       
    basicstyle=\ttfamily\small,       
    keywordstyle=\color{violet}\bfseries, 
    commentstyle=\color{green!50!black}\itshape, 
    stringstyle=\color{red},          
    numbers=left,                     
    numberstyle=\tiny\color{gray},    
    stepnumber=1,                     
    numbersep=10pt,                   
    backgroundcolor=\color{gray!5},   
    rulecolor=\color{black},          
    frame=shadowbox,                  
    tabsize=4,                        
    captionpos=b,                     
    breaklines=true,                  
    breakatwhitespace=true,           
    showspaces=false,                 
    showtabs=false,                   
    showstringspaces=false,           
    morekeywords={li, sw, lb, csrr, csrc, csrs, addi}, 
}

\begin{document}

\title{GoldenFuzz: \\Generative Golden Reference Hardware Fuzzing}
\pagestyle{plain} 
\setcounter{page}{1}

\makeatletter
\newcommand{\linebreakand}{%
  \end{@IEEEauthorhalign}
  \hfill\mbox{}\par
  \mbox{}\hfill\begin{@IEEEauthorhalign}
}
\makeatother

\author{\IEEEauthorblockN{Lichao Wu}
	\IEEEauthorblockA{Technical University of Darmstadt\\
		lichao.wu@trust.tu-darmstadt.de}
	\and
	\IEEEauthorblockN{Mohamadreza Rostami}
	\IEEEauthorblockA{Technical University of Darmstadt\\
		mohamadreza.rostami@trust.tu-darmstadt.de}
	\and
	\IEEEauthorblockN{Huimin Li}
	\IEEEauthorblockA{Technical University of Darmstadt\\
		huimin.li@trust.tu-darmstadt.de}    
	\linebreakand
	\IEEEauthorblockN{Nikhilesh Singh}
	\IEEEauthorblockA{Technical University of Darmstadt\\
		nikhilesh.singh@trust.tu-darmstadt.de} 
	\and
	\IEEEauthorblockN{Ahmad-Reza Sadeghi}
	\IEEEauthorblockA{Technical University of Darmstadt\\
		ahmad.sadeghi@trust.informatik.tu-darmstadt.de}
        }

\IEEEoverridecommandlockouts
\makeatletter\def\@IEEEpubidpullup{6.5\baselineskip}\makeatother
\IEEEpubid{\parbox{\columnwidth}{
		Network and Distributed System Security (NDSS) Symposium 2026\\
		23 - 27 February 2026 , San Diego, CA, USA\\
		ISBN 979-8-9919276-8-0\\  
		https://dx.doi.org/10.14722/ndss.2026.231663\\
		www.ndss-symposium.org
}
\hspace{\columnsep}\makebox[\columnwidth]{}}

\maketitle


\begin{abstract}
Modern hardware systems, driven by demands for high performance and application-specific functionality, have grown increasingly complex, introducing large surfaces for bugs and security-critical vulnerabilities. Fuzzing has emerged as a scalable solution for discovering such flaws. Yet, existing hardware fuzzers suffer from limited semantic awareness, inefficient test refinement, and high computational overhead due to reliance on slow device simulation.

In this paper, we present GoldenFuzz, a novel two-stage hardware fuzzing framework that partially decouples test case refinement from coverage and vulnerability exploration. GoldenFuzz leverages a fast, ISA-compliant Golden Reference Model (GRM) as a ``digital twin'' of the Device Under Test (DUT). It fuzzes the GRM first, enabling rapid, low-cost test case refinement, accelerating deep architectural exploration and vulnerability discovery on DUT. During the fuzzing pipeline, GoldenFuzz iteratively constructs test cases by concatenating carefully chosen instruction blocks that balance the subtle inter- and intra-instructions quality. A feedback-driven mechanism leveraging insights from both high- and low-coverage samples further enhances GoldenFuzz's capability in hardware state exploration. Our evaluation of three RISC-V processors, RocketChip, BOOM, and CVA6, demonstrates that GoldenFuzz significantly outperforms existing fuzzers in achieving the highest coverage with minimal test case length and computational overhead. GoldenFuzz uncovers all known vulnerabilities and discovers five new ones, four of which are classified as highly severe with CVSS v3 severity scores exceeding seven out of ten. It also identifies two previously unknown vulnerabilities in the commercial BA51-H core extension.
\end{abstract}


%
\IEEEpeerreviewmaketitle

\section{Introduction}
Modern processors contain billions of transistors, multiple cores, and sophisticated performance optimization mechanisms that offer unprecedented computational capabilities. This complexity simultaneously enlarges the surface of the hardware attack, exposing systems to a broader range of functional bugs and security-critical flaws, for example, system instability~\cite{IntelUnstable}, incorrect computation~\cite{Pentium,TLBBug}, and information leakages~\cite{Borrello:2022:AEPIC,kocher2020spectre,lipp2018meltdown}. Fixing such flaws post-fabrication could be a challenging and costly task~\cite{Intel}. Consequently, detecting and mitigating hardware vulnerabilities \nadd{in the pre-silicon stage} becomes crucial to preserving system stability and security with a reduced budget.

The hardware community has proposed several \nadd{security} verification techniques, from formal static methods to simulation-based dynamic approaches~\cite{sarangi2006phoenix,deutschbein2018mining,wile2005comprehensive,dessouky2019hardfails,clarke2011model,wagner2007engineering,hicks2015specs, li2011caisson,li2014sapper,zhang2015hardware}.
In recent years, fuzzing has gained attention for its capacity to automate vulnerability detection and scalability on complex hardware designs. 
By systematically probing target devices, hardware fuzzing excels in discovering bugs and security vulnerabilities, making it a practical tool to improve hardware security in increasingly sophisticated computing systems~\cite{trippel2022fuzzing, kande2022thehuzz,chen2023hypfuzz,rostami2024beyond,rostami2024specure,rostami2024fuzzerfly,solt2024cascade,hur2021difuzzrtl,xu2023morfuzz,borkar2024whisperfuzz,canakci2023processorfuzz}. Industry leaders, including Intel and Google, are actively investing in hardware fuzzing research~\cite{trippel2022fuzzing} to strengthen their verification efforts. 
Unfortunately, existing fuzzers face several significant challenges. First, they primarily rely on random mutations or heuristics to generate test cases, failing to capture the complex dependencies and execution semantics inherent to modern Instruction Set Architectures (ISAs). This shortcoming leads to duplicated test cases and subpar exploration of edge cases. Second, previous work typically refines test cases in an ad-hoc manner, lacking a structured mechanism to refine input based on semantic insights. Consequently, they struggle to produce meaningful instruction sequences that reach deeper hardware states. 
Although some fuzzers attempt to improve test selection through, for example, steered control flow~\cite{solt2024cascade} or reinforcement learning~\cite{rostami2024beyond}, they often operate without a clear separation between \emph{test case refinement} and \emph{coverage exploration}, leading to inefficiencies in both processes. Finally, existing work relies on executing test cases solely on the slow simulated DUT. Each iteration requires evaluating numerous test cases, leading to excessive computational overhead and limiting the fuzzer’s ability to explore a broad range of hardware states in a constrained time frame. 

\noindent\\
\textbf{Our goals and contributions:} 
We present \ourname, a novel hardware fuzzing framework that addresses these bottlenecks by decoupling the traditional monolithic fuzzing process from direct, continuous interaction with the target hardware. 
\nadd{Concretely, \ourname restructures hardware fuzzing into two distinct stages: the \emph{Golden Reference Model (GRM) fuzzing} and the \emph{Device Under Test (DUT) fuzzing}. In the first fuzzing stage, inspired by the concept of a digital twin, \ourname introduces a fast, software-based GRM of the DUT\footnote{Although the GRM and DUT implementations may differ internally due to hardware-specific optimizations or undocumented behaviors, they both strictly follow the ISA specification.}.} 
This enables rapid exploration and efficient refinement of fuzzing strategies, guiding the generation of test cases that inherently achieve deeper architectural coverage. Subsequently, the optimized fuzzing policy is transferred directly to the DUT fuzzing stage, greatly accelerating the discovery of meaningful hardware behaviors and potential vulnerabilities. \ourname integrates a customized language model that captures the semantic structures of instruction sequences, generating smarter, more targeted test cases, effectively serving as a translator fluent in the processor's intricate architectural dialect. By continuously analyzing both successful and unsuccessful test cases, \ourname adapts its strategy to systematically target complex, hard-to-reach processor states. 
Together, these design innovations enable \ourname to significantly surpass state-of-the-art fuzzers in coverage growth, uncover deeper and previously hidden vulnerabilities, and dramatically reduce computational overhead in hardware fuzzing.
Concretely, we make the following contributions: 
\begin{itemize}
\item We introduce a two-stage hardware fuzzer, \ourname, that first emphasizes rapid test case refinement at low computational cost, then shifts to focused vulnerability detection to uncover critical weaknesses. During the first phase, \ourname, for the first time, utilizes a GRM, a software model designed to execute RISC-V instructions in strict adherence to the ISA specification, to represent the DUT. This implementation facilitates the evolution of test cases to contain richer semantics, potentially achieving higher hardware state coverage. This insight transfers seamlessly to the DUT, reducing discrepancies and maximizing meaningful fuzzing outcomes.
\item We introduce a block-wise test case generation scheme that produces multiple instruction blocks in each iteration and appends the chosen block to subsequent iterations. A new scoring mechanism, integrating inter- and intra-test case evaluations, drives extensive instruction space exploration, ensuring that each progressive block effectively aligns with fuzzing objectives. 
\item Our framework leverages a novel language model–based generator capable of accurately producing assembly instructions by understanding inter-instruction semantics. This generator integrates the target's feedback, enhancing the test case generation capability with higher coverage. Furthermore, for the first time, we incorporate both ``winning'' and ``losing'' test cases (in terms of hardware coverage) into the fuzzing process. By directly comparing these paired cases, our method continually refines its test generation strategy to expose new vulnerabilities.
\item To the best of our knowledge, \ourname identifies all previously known bugs and vulnerabilities. Additionally, \newbugcount new vulnerabilities are discovered in tested cores (RocketChip~\cite{rc}, Boom~\cite{boom}, and CVA6~\cite{cva6}), including four critical ones with CVSS 3.0 scores above seven (from a maximum of ten). \nadd{For the real-world application, it identifies two previously unknown vulnerabilities in the commercial BA51-H core extension.}
\end{itemize}

The remainder of this paper is structured as follows. We provide the necessary background information in Section~\ref{sec:preliminaries}. In Section~\ref{sec:framework}, we describe the design of \ourname in detail; the implementation is introduced in Section~\ref{sec:Implementation}. Section~\ref{sec:evaluation} evaluates the performance of \ourname regarding the hardware coverage and computational cost. Section~\ref{sec:Fuzzing Findings} details the fuzzing findings, including detected vulnerabilities and bugs. Section~\ref{sec:Ablation Study} performs an ablation and hyperparameter study on several critical fuzzing settings. Section~\ref{sec:discussion} discusses the \ourname with more insights. Section~\ref{sec:related} discusses related works. Section~\ref{sec:conclusions} concludes this work.
\section{Preliminaries}
\label{sec:preliminaries}
\subsection{Fuzzing}
\label{subsec:fuzzing_preliminary}

Fuzzing is a widely used methodology for testing and verifying complex hardware and software designs, such as processors, cryptographic modules, and communication protocols, to uncover potential vulnerabilities~\cite{AFL,chen2023hypfuzz,ammann2024dy,chen2019exploring}. 
A traditional fuzzer begins with randomly produced yet valid test stimuli. State-of-the-art fuzzers have employed iterative mutation algorithms to expand the DUT’s state space coverage. Throughout this process, the DUT’s outputs, including execution traces and any crash information, are captured and analyzed. In software contexts, suspicious crashes may directly reveal exploitable bugs. For hardware, the DUT’s execution traces are compared against a GRM or predefined assertions. Any detected discrepancies become red flags for potential vulnerabilities. This iterative cycle repeats until an acceptable level of state-space coverage is achieved or critical weaknesses are discovered. GRM is only used for vulnerability detection; it does not influence the fuzzing pipeline. 

Hardware fuzzing strategies are generally categorized into black, grey, and white boxes based on the degree of internal knowledge available about the DUT~\cite{saravanan2024fuzz,rostami2024fuzzerfly}. 
Among these techniques, coverage-based white-box fuzzing has become particularly popular for hardware verification, as it systematically evaluates state-space exploration using coverage metrics such as finite-state machine (FSM) coverage, line coverage, condition coverage, and multiplexer (MUX) toggle coverage. Within a hardware fuzzing pipeline, initial input seeds are generated and mutated to produce multiple test cases. Feedback from simulation-based coverage analysis of these test cases then guides the selective refinement of promising inputs while discarding unproductive ones. This feedback-driven loop supports efficient navigation of the DUT’s state space; any anomalies or unexpected DUT responses are recorded for follow-up vulnerability assessment.

\subsection{Language Model and Fine-Tuning}
Language models are one of the most advanced methods in Natural Language Processing (NLP), as they significantly increase the capability of intelligent systems to analyze and generate coherent text. They support various tasks by predicting how words naturally follow one another, including translation, summarization, and conversational agents~\cite{zhao2023survey}. These language models often rely on transformer architectures~\cite{vaswani2017attention}, which use attention mechanisms to capture complex relationships across long text sequences. Among the most influential transformer-based models are Generative Pre-trained Transformers (GPT)~\cite{radford2018improving}, built using a unidirectional approach. GPT models predict each successive token using only previous tokens, relying on stacked transformer decoder layers with multi-head self-attention and feed-forward networks.

Training a language model can be highly resource-intensive; fine-tuning becomes an effective and low-cost learning strategy when customizing pre-trained language models for specific tasks. 
Common methods include Reinforcement Learning from Human Feedback (RLHF)~\cite{ouyang2022training}, which aligns the model’s outputs with human preferences or guidelines, and Direct Preference Optimization (DPO)~\cite{rafailov2024direct,meng2024simpo}, which avoids explicit reward shaping in favor of pairwise comparisons between candidate outputs. These methods enable the model to capture subtle human-defined quality metrics and to produce responses that align with particular application needs.
\section{\ourname}
\label{sec:framework}
\subsection{General Framework}
\label{subsec:exploration and exploitation}
An overview of \ourname is shown in Fig.~\ref{fig:general_framework}. First, our fuzzer, powered by a customized language model (LM), is pre-trained on a corpus of assembly instructions. Next, instead of immediately testing on the slower DUT, the LM first interacts with a fast software-based GRM. During this stage, the LM generates small groups of instructions, denoted as \emph{instruction blocks}, based on previously tested examples. These instruction blocks are rapidly evaluated on the GRM, and the feedback is used to teach the LM to generate more valid and meaningful instruction sequences.

After sufficient refinement, the DUT fuzzing stage starts, where the improved LM now targets the actual hardware (DUT) with a similar fuzzing pipeline. Because the LM was previously \nadd{refined} to produce instruction blocks with high semantic validity, test cases at this stage have a greater potential to explore deep, complex hardware states, achieving more effective testing while minimizing slow hardware simulation overhead. Finally, by comparing execution traces from the DUT against those from the GRM, \ourname efficiently identifies discrepancies that reveal vulnerabilities and bugs.

\begin{figure}[t]
\centerline{\includegraphics[width=\linewidth]{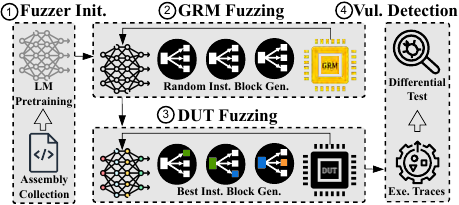}}
\caption{An Overview of the \ourname framework.}
\label{fig:general_framework}
\end{figure}

\subsection{The Motivation Behind the GRM Fuzzing}
\label{subsec:moving from dut to grm}
Traditionally, GRM serves primarily as an oracle to identify mismatches and bugs by comparing DUT outputs against expected behavior. In contrast, \nadd{\ourname applies GRM to coarsely refine the fuzzing policy before the actual DUT fuzzing}, addressing two critical limitations of prior fuzzers~\cite{rostami2024fuzzerfly,kande2022thehuzz,rostami2024beyond}: their tendency to generate test cases containing limited (fewer than 10) executable instructions, and their computational inefficiency due to the prohibitive overhead of cycle-level hardware simulation.

First, since the GRM strictly follows the RISC-V ISA specification, refining the fuzzing policy using GRM feedback encourages the generation of high-quality test cases. \nadd{While acknowledging that subtle deviations or undocumented behaviors in the DUT might not always be perfectly reflected in the GRM, GoldenFuzz addresses this potential mismatch not by attempting to improve coverage directly from the GRM, but by setting a more general objective: to enhance test case validity based on the ISA. This approach enables the fuzzer to learn valid concatenations between instructions and leave the more dedicated, coverage-guided fuzzing to the DUT. Second, the GRM significantly reduces computational overhead. Unlike DUT simulations, the GRM provides a fast, ISA-compliant software environment, enabling efficient fuzzing-policy training. By quickly learning patterns of structurally valid instruction sequences, \ourname transitions smoothly to real hardware, generating high-quality test cases without costly online DUT-guided test case refinement. As empirically validated in Section~\ref{sec:Ablation Study}, this approach facilitates more valid and executable instructions per test case. Besides, as demonstrated in Section~\ref{sec:Fuzzing Findings}, our GRM-driven approach enables the discovery of vulnerability classes that require deep and sustained execution paths, issues that are typically missed by conventional random or shallow test generation strategies.}

\subsection{Fuzzer Initialization}
\label{sec:fuzzer_init}
Before the two fuzzing stages, \ourname is initialized to gain essential instruction generation capability.
\ourname treats a test case, containing a sequence of instructions, as a linguistic construct akin to a \emph{sentence} composed of meaningful \emph{words}. Essentially, each instruction is viewed as a linguistic unit within a higher-level syntax, ensuring its collective arrangement conveys a coherent intent to the DUT. Our approach involves a fuzzer driven by a customized language model (implementation detailed in Section~\ref{subsec:fuzzer}), leveraging its adeptness in natural language processing tasks to generate hardware-focused instruction synthesis. Note that an effective instruction generation engine must internalize both \emph{intra-instruction semantics} (valid and meaningful construction of a single instruction) and \emph{inter-instruction semantics} (synergistic assembly of multiple instructions to unveil vulnerabilities). In this section, we initialize the model with robust intra-instruction knowledge. The refinement of its inter-instruction capabilities through dynamic interaction with the target is elaborated in Section~\ref{subsec:pipeline}.
    
We prepare a diverse set of randomly sampled instructions to equip the fuzzer with a foundational grasp of intra-instruction semantics. 
More concretely, we assemble a corpus of $J$ assembly instructions $\{ I_1, I_2, \ldots, I_J \}$.
The choice of assembly instructions enhances the semantic connections between instructions, facilitating the assembly of meaningful sequences that can probe deeper into vulnerabilities. The instructions corpus is concatenated into a single linear structure $\mathcal{D}$, separated by \texttt{<eoi>} (end of instruction):
\begin{equation}
\mathcal{D} = I_1 \ \texttt{<eoi>} \ I_2 \ \texttt{<eoi>} \ I_3 \ \texttt{<eoi>} \ \cdots \ I_J\ \texttt{<eoi>}.
\end{equation}

This explicit segmentation aids in clarifying instruction boundaries and preserving per-instruction semantics. Next, each instruction $I_i$ is tokenized as $T_i$, forming $\mathcal{T} = T_1 \ T_{\texttt{<eoi>}} \ T_2 \ T_{\texttt{<eoi>}} \ \cdots \ T_J\ T_{\texttt{<eoi>}}$ such that each $T_i$ contains $k_i$ tokens $\mathbf{t}_{i,1}, \mathbf{t}_{i,2}, \ldots, \mathbf{t}_{i,k_i}$. 

We define an auto-regressive objective that compels the model to predict the subsequent token given all previously observed tokens. Let $\mathbf{t} = \{\mathbf{t}_1, \mathbf{t}_2, \ldots, \mathbf{t}_K\}$ represents the entire token sequence of $\mathcal{T}$, with $K (\geq J)$ being the total number of tokens. The model is trained to estimate $P_{\theta}(\mathbf{t}_{i+1} \mid \mathbf{t}_1, \mathbf{t}_2, \ldots, \mathbf{t}_i)$ for each $i$, where $\theta$ denotes the trainable parameters. We minimize the negative log-likelihood $\mathcal{L}$:
\begin{equation}
\mathcal{L}(\theta) = - \sum_{i=1}^{K-1} \log P_{\theta}(\mathbf{t}_{i+1} \mid \mathbf{t}_1, \mathbf{t}_2, \ldots, \mathbf{t}_i).
\label{eq:nll}
\end{equation}

Eq.~\ref{eq:nll} ensures that the model incrementally refines its parameters to produce tokens following previous contexts. 

\subsection{Fuzzing Pipeline}
\label{subsec:pipeline}
As mentioned in Section~\ref{subsec:exploration and exploitation}, both the GRM fuzzing and DUT fuzzing phases in \ourname follow a similar fuzzing pipeline. The key difference lies in their targets and scoring functions. An overview of the \ourname pipeline is shown in Fig.~\ref{fig:fuzzing_pipeline}, consisting of three major steps.
\begin{itemize} 
\item \emph{Test Case Generation and Simulation.} Guided by the current fuzzing policy, we produce test cases by sampling from the \ourname's memory and iteratively creating, selecting (presented by the heart symbol), and concatenating \emph{instruction blocks} (IBs) with each containing multiple instructions. The chosen target (GRM or DUT) executes these IB concatenations and provides feedback. 
\item \emph{Test Case Scoring.} \ourname evaluates IBs differently for each fuzzing stage. For GRM fuzzing, the IB is simply evaluated by its validity. Instead of enforcing validity at the instruction level, an IB that is executable with no GRM exception is considered valid. IB-level validity allows the fuzzer to learn the valid concatenation between instructions, potentially construct multi-IB test cases with non-trivial control flow patterns. For DUT fuzzing, we employ a dual-layer scoring system. This approach incentivizes newly uncovered coverage within a single test case (intra-test scoring) while deducting points for coverage already identified by other tests (inter-test scoring). As a result, this technique drives the fuzzing process toward more hidden hardware states.
\item \emph{Fuzzing Policy Refinement and Memory Update.} Using the computed scores, we form \emph{preference pairs} by identifying “winning” (e.g., \heartcheck, the color indicates the fuzzing iteration) and “losing” (\prohibitsign) test cases based on their scores. These pairs are directly fed back to the fuzzer, driving it toward producing instruction sequences that deliver improved coverage. At the same time, the fuzzer’s memory is updated to maintain a consistent record of test cases aligned with the evolving fuzzing policy.
\end{itemize}
\begin{figure*}[t]
\centerline{\includegraphics[width=\linewidth]{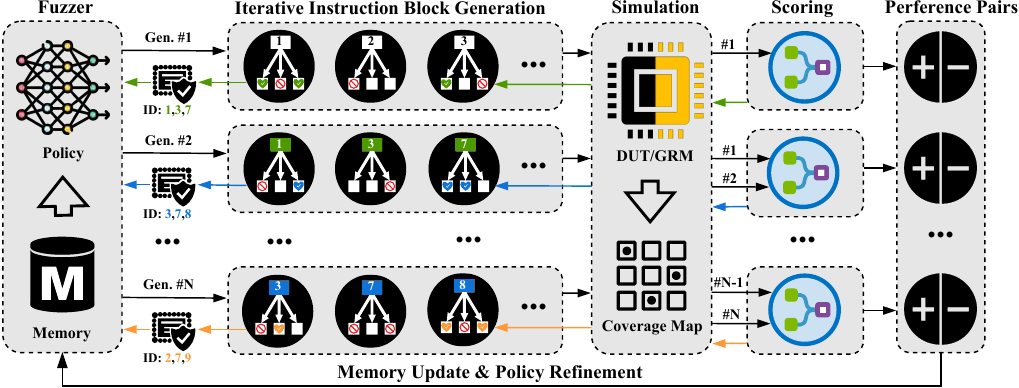}}
\caption{An Overview of the \ourname Pipeline.}
\label{fig:fuzzing_pipeline}
\end{figure*}
\subsubsection{Block-Wise Test Case Generation}
\label{subsubsec:Block-Wise Test Case Generation}

One of the core novelties in \ourname is a structured, block-wise test case generation strategy that balances complexity, coverage, and learning efficiency. Rather than generating full instruction sequences in a single step, \ourname constructs test cases incrementally from smaller units called instruction blocks (IBs). Each IB contains one or more instructions, depending on the current fuzzing configuration. In each fuzzing iteration $i$, \ourname selects a set of previously generated IBs either randomly or based on performance, depending on the fuzzing stage (see Fig.~\ref{fig:general_framework}). These selected IBs are used as prefixes to generate new candidate blocks. Precisely, let $b_i^j$ denote the $j$-th selected block from iteration $i$. For each $b_i^j$, we generate $N$ new instruction blocks, forming a candidate set $\mathcal{B}_{i+1} = \{b_{i+1}^1, b_{i+1}^2, \ldots, b_{i+1}^N\}$. Each new test case is then constructed by concatenating a selected prefix $b_i^j$ with one of the newly generated candidates $b_{i+1}^k \in \mathcal{B}_{i+1}$, resulting in $b_{i,i+1} = b_i^j \oplus b_{i+1}^k$ (for simplicity, we omit the specific $j$ and $k$ in the notation). This strategy has three key advantages. First, by using $b_i^j$ as a starting point, we anchor the execution in a known hardware state, allowing the fuzzer to focus its exploration on how the new block $b_{i+1}^k$ affects the state transition. Second, this block-wise construction simplifies learning: instead of reasoning over full test cases of length $N$ instructions, the fuzzer only needs to learn over smaller blocks of size roughly $N/M$ when the test case is divided into $M$ blocks. This decomposition reduces the complexity of the learning problem and improves convergence during feedback-guided optimization. Finally, repeatedly sampling new IBs allows the fuzzer to explore different (and potentially new) hardware spaces. As a result, each iteration pushes the hardware state exploration further, guiding the search process into less-discovered state space while preserving the insight gained from earlier discoveries.
\begin{algorithm}[!ht]
\footnotesize
\caption{Instruction Block Generation.}
\begin{algorithmic}[1]
\Require $\mathcal{B}_{i}, \texttt{fuzzer}, num\_blocks, num\_inst\_per\_block$ 
\Ensure{$\mathcal{B}_{i+1}$}
\State $\mathcal{B}_{i+1} \gets \{\}$
\For{$b$ \textbf{in} $\mathcal{B}_{i}$}   
\State $cnt \gets 0$
    \While{$cnt < num\_inst\_per\_block$}
        \State $actions \gets \texttt{fuzzer}(b, num\_blocks)$
        \State $b, eoi \gets \texttt{step}(b, actions)$
        \If{$eoi$}
            \State $cnt \gets cnt+1$
        \EndIf
    \EndWhile
    \State $\mathcal{B}_{i+1} \gets \texttt{append}(b)$
\EndFor
\State $\mathcal{B}_{i+1} \gets \texttt{remove\_dead\_blocks}(\mathcal{B}_{i+1})$
\If{GRM\_fuzzing}
    \State $\mathcal{B}_{i+1} \gets \texttt{get\_random\_n}(\mathcal{B}_{i+1})$
\Else
    \State $\mathcal{B}_{i+1} \gets \texttt{get\_best\_n}(\mathcal{B}_{i+1})$
\EndIf
\end{algorithmic}
\label{alg:gen_inst_block}
\end{algorithm}

Algorithm~\ref{alg:gen_inst_block} illustrates the instruction block (IB) generation process for a single fuzzing iteration. The process begins with the selected IB set $\mathcal{B}_{i}$ from iteration $i$. To generate the next set $\mathcal{B}_{i+1}$, the fuzzer extends each IB $b \in \mathcal{B}_{i}$ by taking a sequence of actions (Line 5), each appending a new instruction token to $b$ (Line 6). This continues until a special separator token, \texttt{<eoi>}, is generated, signaling the end of an instruction (Line 7). Once the number of instructions in the current IB reaches the predefined limit, \texttt{num\_inst\_per\_block} (Line 4), the fully constructed block, consisting of the original prefix and the newly generated instructions, is added to $\mathcal{B}_{i+1}$ (Line 11). After generation, each IB in $\mathcal{B}_{i+1}$ is simulated to determine whether it is ``dead''. We consider an IB ``dead'' if it meets one of two conditions: (1) it is syntactically invalid, meaning it violates ISA specification, or (2) it prematurely terminates execution, for example by containing control-flow terminators like a \texttt{ret} (return) instruction. Since dead IBs cannot be meaningfully extended in future iterations, they are excluded from subsequent iterations. However, during the GRM fuzzing stage, these IBs still serve as useful negative (``losing'') examples, contributing to the fuzzing policy refinement.

\subsubsection{Inter and Intra-test Case Scoring}
\label{subsubsec:inter and intra-test case scoring}
Traditional hardware fuzzing methods usually assess each test case fully, measuring total coverage only after the test case is complete. In contrast, in the DUT fuzzing stage, \ourname evaluates each IB and employs a dual-layer scoring system:

\noindent
\textbf{Intra-test case scoring.} 
We employ intra-test case scoring to incentivize newly uncovered coverage points within a single test case.
Let \(b_{i}\) be the \(i\)-th IB in a test case, and let \(\mathcal{H}_{b_{i}}\) represent the set of coverage points already revealed by \(b_{i}\). When transitioning to IB \(b_{i+1}\), the coverage newly discovered by combining \(b_{i}\) with \(b_{i+1}\) is denoted by \(\mathcal{G}(b_{i,i+1})\). As newly uncovered states are more valuable (more likely to trigger unexplored hardware stats), they have a higher weight $(w)$ than previously seen states. Specifically, we use two base weights, \(\alpha\) and \(\beta\) (with \(\alpha < \beta\)), so that
\begin{equation}
w(x) = 
\begin{cases}
\alpha, & \text{if } x \in \mathcal{H}_{b_{i}},\\
\beta,  & \text{if } x \notin \mathcal{H}_{b_{i}},
\end{cases}
\label{eq:intra-test case scoring}
\end{equation} 
where $x$ denotes a coverage point. \(\alpha\) is a lower reward for coverage points that merely repeat what was already observed by previous IBs, whereas \(\beta\) is a higher reward to incentivize new coverage. 

\noindent
\textbf{Inter-test case scoring.} 
Inter-test case scoring deducts the coverage score already identified by other test cases.
Let \(f(x)\) be the frequency with which a coverage point \(x\) is encountered by test cases other than the current one. We introduce a small constant factor to reduce \(\beta\) in proportion to \(f(x)\), thereby lowering the reward for states that are already popular among different test cases:
\begin{equation}
\beta'(x) = \max\bigl(\beta - f(x)\cdot \text{factor},\ \alpha\bigr).
\label{eq:beta'}
\end{equation} 

Intuitively, as \(f(x)\) increases, \(\beta'(x)\) linearly transitions from \(\beta\) (fully rewarding) down toward \(\alpha\), reflecting the diminishing scores assigned to repeatedly covered points.

Combining two scoring schemes, the overall score for the transition from \(b_{i}\) to \(b_{i+1}\) is defined by summing the adjusted weights for every coverage point \(x\) in \(\mathcal{G}(b_{i,i+1})\). We have:
\begin{equation}
\text{S}(b_{i,i+1}) = \sum_{x \in \mathcal{G}(b_{i,i+1})} w'(x),
\label{eq:score}
\end{equation} 
where $\beta$ in Eq.~\ref{eq:intra-test case scoring} is replaced with $\beta'(x)$ in Eq.~\ref{eq:beta'} to form $w'(x)$. \nadd{This strategy enables dynamic score adjustment in response to coverage frequency, potentially mitigating the risk of the fuzzer over-prioritizing already successful test cases. As shown in Section~\ref{sec:Coverage Evaluation}, the hardware coverage keeps increasing with more test cases, while other fuzzer are saturated early.}

\subsubsection{Fuzzing Policy Refinement}
\label{subsubsec:fuzzing policy refinement}
After scoring test cases, the next step is to refine the fuzzing policy using a principled, data-driven approach. Traditional hardware fuzzers, as described in Section~\ref{subsec:fuzzing_preliminary}, typically adopt a mutation-based strategy inspired by the American Fuzzy Lop (AFL) pipeline~\cite{AFL}. They randomly mutate successful test cases while discarding unsuccessful ones. While straightforward, this approach is inefficient and wastes valuable information: it neglects insights from “losing” test cases that could guide future iterations and makes refining “good” test cases erratic, as it lacks feedback on which mutations are most effective.

\ourname overcomes these limitations by directly integrating feedback into the fuzzing process. Instead of randomly guessing which test cases are promising, \ourname explicitly pairs “winning” and “losing” test cases based on their scores, forming \emph{preference pairs}. These pairs allow the fuzzer to efficiently utilize feedback regardless of test case coverage and refine its preference (i.e., fuzzer's fuzzing policy) through direct comparison. By consistently prioritizing “winning” test cases, the fuzzer evolves toward generating test cases with higher coverage, thus probing into the deeper state space. 

Concretely, we define the intrinsic reward of a test case $b$ using the model’s likelihood of generating it: 
\begin{equation} 
r(b) = \frac{\beta}{|b|} \sum_{i=1}^{|b|} \log \pi_\theta(t_i \mid t_{<i}), 
\label{eq:prob_reward}
\end{equation} 
where $t_i$ denotes the $i$-th token of $b$; $t_{<i}$ represents the sequence of tokens preceding $t_i$; $\pi_\theta$ is the current fuzzing policy parameterized by $\theta$; and $\beta$ is introduced to scale the reward. This per-token likelihood quantifies how “natural” or policy-consistent a given test case is. Normalizing by sequence length ensures that test cases are not unfairly penalized or rewarded based on their instruction length alone, thus forcing \ourname to generate high-quality IBs instead of just making, e.g., long IBs, to win more rewards. 

Using feedback from the GRM or DUT, pairwise preferences are established by comparing test cases and identifying a “winner” based on the scoring function (Eq.~\ref{eq:score}). Given two test cases $b_w$ (“winner”) and $b_l$ (“loser”), employing Simple Preference Optimization (SimPO)~\cite{meng2024simpo}, a specialized form of Direct Preference Optimization~\cite{rafailov2024direct} with lower computation cost, we update the fuzzing policy by minimizing: 
\begin{equation} 
\mathcal{L}(\theta) = -\mathbb{E}_{(b_w, b_l) \sim \mathcal{B}} \left[ \log \sigma\left( r(b_{i,w}) - r(b_{i,l}) - \gamma \right) \right], 
\label{eq:simpo_loss}
\end{equation}
where $\mathcal{B}$ denotes the set of preference pairs constructed from the GRM evaluations; $\sigma$ denotes the sigmoid function; $\gamma$ establishes a target reward margin, ensuring that preference differences translate into meaningful fuzzing policy shifts. Eq.~\ref{eq:simpo_loss} encourages the fuzzer to shift output probability mass toward generating “winner” instruction sequences that are empirically more likely to uncover new coverage points.

However, naively applying Eq.~\ref{eq:simpo_loss} to our fuzzing scheme introduces significant challenges. DPO is typically employed as an offline optimization method, where preference pairs are collected before optimization. In contrast, hardware fuzzing demands an online optimization approach, where the fuzzing policy is refined iteratively during runtime. This continuous refinement introduces the risk of overfitting. 
Over time, the generated instruction risks drifting toward narrower instruction sets, neglecting other potentially valuable regions of the state space in the “distributional tails”~\cite{shumailov2024ai}. Such rare points (a.k.a. corner cases) are critical for exposing deep and previously unseen vulnerabilities. Eventually, the fuzzer could collapse, unable to produce even syntactically correct instructions.

To address this, we introduce a fuzzing memory $\mathcal{M}$ to balance immediate gains with maintaining exploration diversity. After each iteration $i$, we identify $N$ top IBs and preference pairs from $\mathcal{B}_{i}$ and store these “exemplars” in $\mathcal{M}$. $\mathcal{M}$ follows the “first-in-first-out” principle: we remove the oldest set from iteration $i-N$ to prevent unbounded growth and maintain a rolling window of strong candidates: 
\begin{equation} 
\mathcal{M} \leftarrow (\mathcal{M} \setminus \mathcal{B}^{\star}_{i-N}) \cup \mathcal{B}^{\star}_{i+1}. 
\end{equation} 

The IB and preference pairs are sampled from $\mathcal{M}$ during a new refinement iteration or fuzzing policy. We assign exponential recency weighting to each sample so that recent samples are more likely to be sampled, but keep older samples in the mix. 
This prevents forgetting earlier coverage strategies and guards against overfitting to the current iteration’s local maxima~\cite{fedus2020revisiting}. 
Over time, this yields a stable fuzzing process.

\subsection{Bug and Vulnerability Detection}
\label{subsec:dvulndetect}
Differential testing is extensively utilized in hardware fuzzing to identify ``crashes''. Under this methodology, a single test case is executed on both the DUT and a GRM based on the ISA. The execution traces obtained from these models are then compared~\cite{rostami2024fuzzerfly,kande2022thehuzz,rostami2024beyond,solt2024cascade}. In alignment with state-of-the-art hardware fuzzers, \ourname employs differential testing involving the DUT and the GRM. This approach has demonstrated its effectiveness across various hardware fuzzers, particularly in the context of RISC-V fuzzers, which have, to date, facilitated the discovery of most of bugs and vulnerabilities~\cite{rostami2024fuzzerfly,kande2022thehuzz,rostami2024beyond,solt2024cascade,chen2023hypfuzz,hur2021difuzzrtl},

Although highly effective at identifying issues, differential testing can generate many mismatches, many of which are either duplicates or false positives~\cite{rostami2024fuzzerfly}. 
\nadd{Given the manual nature of vulnerability analysis, prolonged investigation of erroneous or redundant mismatches can significantly hinder the efficiency and scalability of hardware fuzzing, particularly as design complexity grows. In our workflow, each unique mismatch, whether classified as a true bug or a false positive, initially requires manual inspection. For the five new vulnerabilities, confirmation and classification typically took between 5 and 30 minutes per case, depending on the complexity of the scenario and the need to trace privilege transitions or instruction sequences. In the early stages of fuzzing, when many mismatches are new, this manual effort could increase.
To (partially) address this limitation and improve scalability, we propose a filtering approach to streamline the analysis process. After each mismatch is analyzed, it is added to a known mismatch list along with its environmental context (e.g., privilege level, instruction type, register values, and exception details). If the same mismatch recurs, the system can automatically classify it without further manual intervention. As fuzzing progresses, the proportion of previously seen mismatches increases, and the filtering mechanism substantially reduces the number of cases requiring manual analysis. This approach ensures that, even for large and complex designs, the manual effort required for vulnerability confirmation and analysis remains manageable.}
\section{Implementation}
\label{sec:Implementation}

\subsection{Fuzzer Design and Pre-training}
\label{subsec:fuzzer}
We implement our fuzzer using a GPT-2 language model tailored for hardware instruction generation. Employing open-source architecture, consisting of 1.5 billion parameters and a vocabulary of over 50\,000 tokens, poses two major challenges. First, GPT-2’s pretrained parameters are grounded in natural (English) language, which differs substantially from assembly and binary code syntax. 
Second, GPT-2 employs Byte-Pair Encoding (BPE) to handle tokenization, splitting infrequently occurring words into smaller subword units. Although BPE benefits broad-domain language tasks, it introduces unnecessary complexity for hardware instructions. 
To address these issues, we built a customized GPT model designed explicitly for the RISC-V instruction set. Instead of relying on subwords, we assign individual tokens to each opcode and operand. This straightforward tokenization strategy preserves opcode/operand-level correctness without vastly increasing the vocabulary size. 
Besides, the token length of instruction is more than half reduced compared with BPE, making the training stage much easier, as the model does not need to learn long token patterns. 
During our preliminary study, we also experimented with reducing the model size by decreasing the number of layers, attention heads, and embedding dimensions. However, this led to suboptimal performance during fuzzing, potentially due to the reduced model capacity limiting its ability to capture the complex dependencies and semantics required to generate effective and diverse instruction blocks.

\nadd{Training data preparation is fundamental to the fuzzer's performance and its capacity to explore the DUT. To ensure comprehensive test case coverage with different instructions, our custom fuzzer model is trained from scratch using 10 million randomly generated RISC-V assembly instructions, including all possible RISC-V instructions and extensions: the fundamental 32-bit and 64-bit RISC-V ISAs and various extensions, such as integer multiplication/division, single-precision floating-point operations, atomic memory operations, compressed instructions, and machine-level instructions. This exhaustive inclusion allows our fuzzer to generate relevant test cases with maximum diversity, including less frequently utilized or specialized components of the instruction set. Besides, due to the generative nature of the underlying LLM, diverse and even syntactically incorrect instructions still emerge naturally. While undesirable in typical LLM tasks, this characteristic is beneficial in hardware fuzzing, as such incorrectness helps evaluate corner cases.}
The initial training, described in Section~\ref{sec:fuzzer_init}, involves 50,000 epochs with a learning rate 1e-6 and takes approximately one hour on one NVIDIA A6000 GPU. During the subsequent fuzzing policy refinement phase (discussed in Section~\ref{subsec:fuzzing process}), we lower the learning rate to 2e-7 to maintain stable fine-tuning and ensure that the fuzzer’s output remains coherent.

\subsection{Hardware Fuzzing Settings}
\label{subsec:fuzzing process}
Recall in Section~\ref{subsec:pipeline}, \ourname initiates the fuzzing process by generating test cases constructed from multiple instruction blocks (IBs). In our configuration, each test case contains five IBs, each with six instructions. Preliminary experiments confirm that carefully chosen sets of around 30 instructions are generally sufficient to detect vulnerabilities, aligning with our observations on state-of-the-art fuzzers. We assess this hyperparameter choice in Section~\ref{subsec:The Selection of Instruction Blocks Settings}. In both fuzzing stages, 80 IBs are sampled from the fuzzing memory in each iteration, each of which forms the input to the language model for generating five new IBs.

Each generated test case is executed on the GRM or the DUT. During the GRM fuzzing, as mentioned in Section~\ref{subsec:moving from dut to grm}, the fuzzing policy was refined by the test case validity judged by the GRM. 
In the DUT fuzzing stage, hardware coverage, through Synopsys VCS~\cite{vcs}, is used as the feedback, which is determined by examining a range of metrics that capture design behavior. These include finite state machine (FSM) coverage, condition coverage, and line coverage. FSM coverage evaluates how thoroughly the test cases explore the different states and transitions within state machines. Condition coverage evaluates if logical conditions, such as branches and conditional expressions, have been covered, thus revealing how well decision points in the logic are tested. Line coverage measures how many lines of the Register Transfer Level (RTL) code have been stimulated. These coverage metrics offer a comprehensive assessment of each test case’s effectiveness in validating the hardware, guiding the fuzzing process toward more complete and revealing explorations of the design’s behavior. Besides, we randomize initial register values to maximize the likelihood of triggering corner cases. In terms of scoring function, we choose $\alpha = 0.1$, $\beta = 1$, and $\text{factor} = \text{1e-5}$ to heavily reward the exploration of new hardware states.

After computing scores, we form preference pairs of “winning” and “losing” IBs to refine the fuzzing policy, guided by the test case score using coverage metrics as inputs. We consider the best-score and worst-score IBs to be “winners” and “losers”, respectively. Hyperparameter tuning is crucial to this iterative learning process. We pay special attention to three hyperparameters: the learning rate, the reward scaling factor $\beta$ for sequence likelihood (Eq.~\ref{eq:prob_reward}), and the target reward margin $\gamma$ (Eq.~\ref{eq:simpo_loss}). We test a range of values (1e-7, 2e-7, 5e-7, 1e-6, and 5e-6) and find that a smaller learning rate (e.g., 2e-7) provides efficient refinements and prevents the fuzzer from collapsing into incoherent or repetitive outputs. 
Similarly, $\beta = 10$ (chosen from a range of 1 to 10) delivers a balanced scaling between winning and losing responses. For the target reward margin $\gamma$, we settle on 0.8 after a fine-grained grid search from 0.1 to 1 in increments of 0.1. Each training iteration uses a batch size of 128, enabling efficient GPU memory usage and stable fuzzing policy update. \nadd{In general, these hyperparameters can be directly applied to new targets, with adaptation (if needed) starting from small learning rates and gradual adjustment of reward scaling.}

\ourname employs the Spike simulator~\cite{spike} as the GRM during the profiling stage, while DUT implementations include RocketChip~\cite{rc}, BOOM~\cite{boom}, and CVA6~\cite{cva6}. We rely on a combined simulation workflow involving Synopsys VCS and Spike for vulnerability detection. Synopsys VCS provides detailed hardware simulation traces, recording register updates and memory operations at each executed instruction’s boundary. Concurrently, Spike serves as the high-level reference model for RISC-V ISA execution, producing idealized reference traces that outline the expected register and memory states after each instruction when running a RISC-V binary. Our framework identifies discrepancies in register values, memory addresses, and memory contents by comparing the hardware simulation traces from VCS and Spike’s execution traces utilizing a mismatch detector. Any mismatch flags a potential bug or vulnerability in the DUT or Spike, which will be analyzed manually by \ourname's user to confirm the bug or vulnerability. 

We developed an automated framework to identify mismatches by parsing trace outputs generated by the target cores. For each test case, this tool processes execution traces for both the core and the GRM, which include details such as time, clock cycles, addresses, instructions, execution privilege levels, register values post-instruction execution, memory transactions, and exception details. These traces are then compared instruction by instruction. Since the initial portions of all test cases are identical due to the initialization of the environment, the parser skips these instructions. Upon detecting a mismatch between the core and GRM traces, the framework applies the filtering approach described in Section~\ref{subsec:dvulndetect}. The mismatch is disregarded if the environmental information associated with the mismatch aligns with any predefined filter criteria. Otherwise, it is logged in a file for further manual analysis.
\section{Performance Evaluation}
\label{sec:evaluation}
This section evaluates the performance of \ourname through a comprehensive analysis and benchmark on hardware coverage and computational cost. 

\subsection{Hardware Coverage}
\label{sec:Coverage Evaluation}
\ourname adopts a coverage-guided white-box fuzzing strategy to maximize exploration of hardware states, thereby elevating the probability of exposing bugs and vulnerabilities. For the hardware coverage benchmark, we compare \ourname against four state-of-the-art fuzzers: Cascade~\cite{solt2024cascade}, DifuzzRTL~\cite{hur2021difuzzrtl}, TheHuzz~\cite{kande2022thehuzz}, and ChatFuzz~\cite{rostami2024beyond}, with a special emphasis on Cascade, the most recent among them, \nadd{and ChatFuzz, the most recent LLM-based fuzzer}. For a thorough assessment, we measure condition, line, and FSM coverage across three RISC-V cores: RocketChip~\cite{rc}, BOOM~\cite{boom}, and CVA6~\cite{cva6}. \nadd{Due to page limit}, we limit our analysis of DifuzzRTL and TheHuzz to condition coverage on RocketChip.
\begin{figure}[htbp]
\centering
\subfloat[\nadd{RocketChip}]{\includegraphics[scale=0.28]{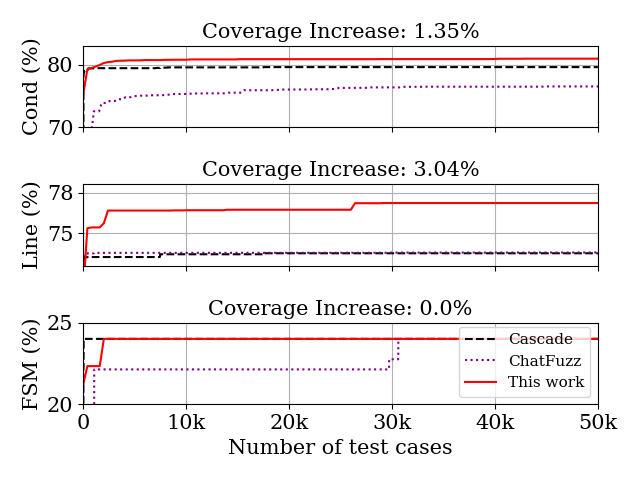}
\label{fig:benchmark_cascade_rocket}}
\subfloat[\nadd{Boom}]{\includegraphics[scale=0.28]{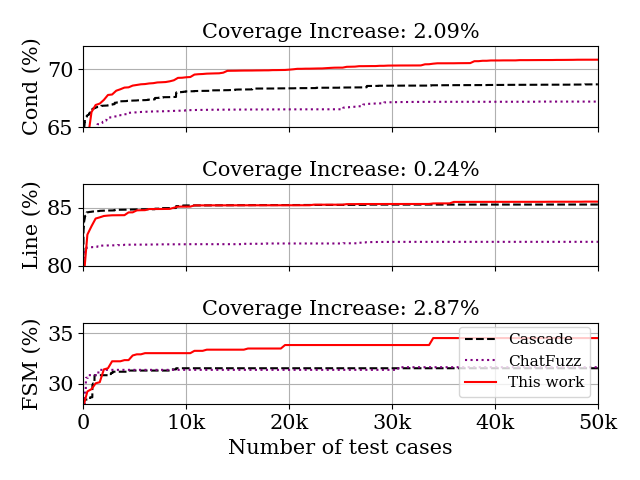}
\label{fig:benchmark_cascade_boom}}
\\
\subfloat[\nadd{CVA6}]{\includegraphics[scale=0.28]{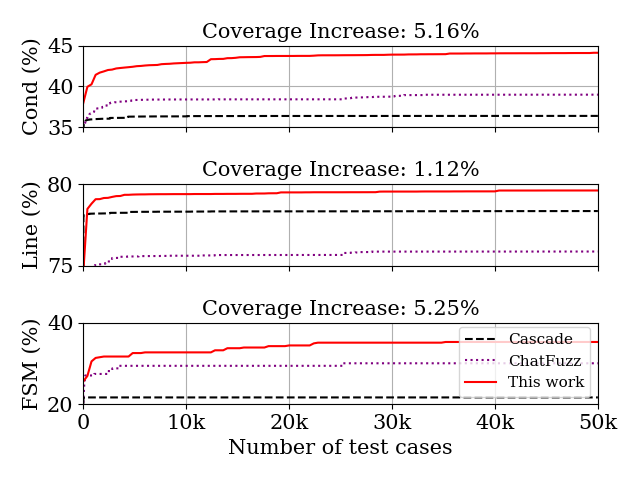}
\label{fig:benchmark_cascade_cva6}}
\subfloat[All Fuzzers on RocketChip]{\includegraphics[scale=0.28]{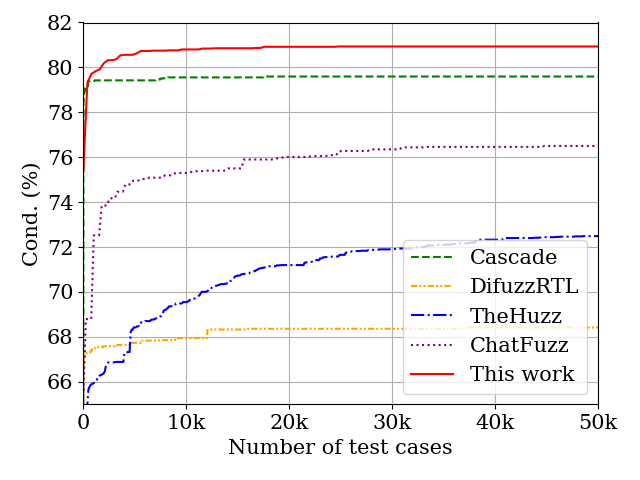}
\label{fig:benchmark_rocket_cond_all}}
\caption{\nadd{Coverage Benchmark.}}
\label{fig:benchmark_cascade}
\end{figure}

The results are shown Fig.~\ref{fig:benchmark_cascade}. We adjust the figure scale to better visualize the trend in coverage. The coverage increase introduced by \ourname\ \nadd{compared with the best performing fuzzer} is shown in the figure title. \ourname consistently outperforms Cascade \nadd{and ChatFuzz} across all tested cores and coverage metrics, except FSM coverage on RocketChip (bottom figure in Fig.~\ref{fig:benchmark_cascade_rocket}), where both fuzzers achieve \nadd{the same coverage}. Notably, Cascade employs basic block concatenation, where each block terminates with an instruction that modifies the program counter. This approach allows its test cases to grow to thousands of instructions, with 10\,000 instructions as optimal for maximizing coverage and vulnerability detection. In contrast, \ourname operates with significantly shorter test cases with 30 instructions. Consequently, Cascade, in some test scenarios, achieves higher coverage at the beginning due to its longer test sequences. However, the early coverage gains do not inherently sustain long-term exploration. 
The coverage of \ourname steadily increases across all three cores, eventually surpassing Cascade, whose coverage plateaus earlier.
\nadd{On the other hand, although ChatFuzz also employs a language model, the generated binary test case constrains the fuzzer to understand the inter- and intra-instruction semantics, eventually leading to lower coverage.} 

Next, we compare coverage among all tested fuzzers on RocketChip using condition coverage metrics, again including Cascade for completeness. As presented in Fig.~\ref{fig:benchmark_rocket_cond_all}, \ourname significantly exceeds the coverage achieved by DifuzzRTL, TheHuzz, and ChatFuzz, each plateaus quickly. By contrast, \ourname maintains superior coverage even as the number of test cases grows. Impressively, it achieves coverage comparable to that of other fuzzers using test cases of only 30 instructions and less than 1\% of test cases, demonstrating its efficiency in exploring diverse hardware states. This advantage translates directly to detecting bugs and vulnerabilities with minimal hardware simulation overhead, affirming \ourname as a high-performance tool for hardware security assessments. We evaluate the robustness of \ourname in Section~\ref{subsec:robustness}.

\subsection{Computational Cost}\label{sec:Computational Cost}
The computational efficiency of a fuzzer directly impacts its overall runtime performance. The overhead in \ourname primarily stems from \nadd{four} components: \nadd{(1) fuzzer pertaining,} (2) test case generation, (3) fuzzer's fuzzing policy refinement, and (4) DUT instrumentation. \nadd{The pre-training of the fuzzers, a one-time task, takes around one hour to finish on an NVIDIA A6000 GPU. Although the three reset components} primarily run on a GPU, we analyze their computational performance on both CPU and GPU for a fair comparison with related works.

On an AMD EPYC 9684X CPU, \ourname generates each test case in just 0.34 seconds, which shrinks to 0.012 seconds on an A6000 GPU. These times significantly outperform advanced fuzzing tools such as Cascade (2.06 seconds per test) and TheHuzz (2.47 seconds) \nadd{in the same CPU setting}. 
Indeed, \ourname’s instruction block-based workflow keeps overhead low: each new test only requires (1) choosing existing instruction blocks and (2) appending one block of six additional instructions. By scaling up parallel input generation, trading off memory for the ability to produce many tests simultaneously. 

In the GRM fuzzing stage, evaluating a test case takes around 0.004 seconds \nadd{with CPU}, contributing only minimal additional cost. Direct preference optimization appears to be time-intensive. However, with our customized and small language model and limited preference pairs per iteration, each tuning iteration completes in under 40 seconds on an A6000 GPU \nadd{and less than 200 seconds using only a CPU}. The bottleneck for \ourname lies in DUT instrumentation, which averages 1.36 seconds per test case when running 80 test cases in parallel. Despite this, \ourname reduces its total testing volume by \emph{more than half} compared to other fuzzers and still achieves comparable or even superior coverage (Fig.~\ref{fig:benchmark_rocket_cond_all}), substantially reducing the overall computational overhead.

\section{Fuzzing Findings}\label{sec:Fuzzing Findings}

\subsection{Testcase Quality}\label{sec:Testcase Quality}
While Section~\ref{sec:Coverage Evaluation} has shown that \ourname significantly improves hardware coverage, this section provides a deeper empirical analysis of the mechanisms driving this improvement. Concretely, we explain why \ourname achieves higher and faster coverage compared to existing fuzzers.

\vspace{2mm}
\begin{lstlisting}[caption={Testcase Quality Running Example}, label={lst:test_quality}, numbersep=5pt,xleftmargin=2pt]
li   x2, 0xa9b1d00fffffffff
csrs pmpaddr0, x2
...
li   x14, 0xff0f0fccdfaaaa1f
csrs pmpcfg0 , x14
...
li   x7, 0x000000000005bcfa
csrs mstatus, x7
...
mret
\end{lstlisting}

Listing~\ref{lst:test_quality} shows a simplified test case generated by \ourname. 
\nadd{While the instruction values and memory operations involved in this testcase may initially seems arbitrary, they actually constitute a minimal and valid sequence required to transition the processor from Machine (M-mode) to Supervisor (S-mode) privilege. Traditional fuzzers and verification tools often fail to uncover such vulnerabilities because they struggle with the enormous search space and typically lack the semantic understanding needed to generate these specific privilege-escalation scenarios. As a result, these tools are unlikely to produce the precise instruction and memory patterns necessary to trigger the vulnerabilities within a reasonable timeframe.
In contrast, \ourname leverages semantic guidance to synthesize such targeted behaviors efficiently, often within the first 1\,000 test cases. This capability explains why \ourname was able to detect the five new vulnerabilities that other tools missed. These results empirically support several key mechanisms behind our framework’s effectiveness:}

\begin{itemize}
    \item \emph{Effectiveness of GRM feedback}.  The GRM steers the generation process toward semantically valid test cases that are less likely to trigger immediate exceptions, thus enabling deeper exploration of the DUT. For instance, instructions like \texttt{li} \texttt{x2}, \texttt{0xa9b1d00fffffffff} and \texttt{csrs} \texttt{pmpcfg0}, \texttt{x14} emerge naturally as the fuzzer explores the input space. 
    \item \emph{Coverage-driven DUT fuzzing}. 
    When running on the DUT, feedback based on hardware coverage metrics (e.g., condition or transition coverage) guides the fuzzer to generate tests that explore unvisited states. Privilege transitions (Line 10), such as \texttt{mret} and \texttt{sret}, require multiple conditions to be satisfied. Their presence in the test case demonstrates that \ourname can learn and exploit such constraints to expand coverage.
    \item \emph{Language model integration for enhanced understanding}. 
    By integrating coverage feedback into the training process, the language model-based fuzzer learns to associate instruction patterns with state transitions. For example, the co-occurrence of \texttt{pmpaddr} and \texttt{pmpcfg} reflects the model’s understanding of PMP (Physical Memory Protection) configurations, indicating that the model captures semantic dependencies across instruction sequences.
\end{itemize}

\subsection{Detected Vulnerabilities}\label{sec:Detected Vulnerabilities and Bugs}
\nadd{\ourname identified five previously unknown vulnerabilities in the open-source cores and two from the commercialized BA51-H core~\cite{pinto2024novel}.} These vulnerabilities have been reported to the respective benchmark developers, \nadd{who confirmed that they were unaware of these issues before our disclosure. We have obtained the following Common Vulnerabilities and Exposures (CVE) entries for these vulnerabilities, \texttt{CVE-2025-45883} and \texttt{CVE-2025-45881}}.
The five vulnerabilities we identified are severe (four have a CVSS 3.0 score exceeding 7) and complex, involving multi-instruction execution paths and triggers from different privilege modes. Besides, the effectiveness of \ourname is not limited to discovering these specific vulnerabilities. During our analysis, we observed that our fuzzer could also trigger bugs and vulnerabilities previously reported but remained unresolved. Furthermore, static analysis revealed that our fuzzer could generate test cases for vulnerabilities already addressed in recent updates~\cite{cva6_fix_1,cva6_fix_2}.
In certain instances, our fuzzer even detected bugs not reported in related fuzzing studies, although the benchmark developers acknowledged being aware of these issues.

The following paragraphs explain each vulnerability and present Proof-of-Concept (PoC) code to demonstrate how these vulnerabilities can be triggered. The PoC code examples are simplified and represent minimal subsets of the test cases generated by the fuzzer.

\noindent\textbf{\texttt{V1 \& V2} Incorrect Endiannes Changes in CVA6 Processors.} 
The RISC-V ISA specification defines the \texttt{MBE}, and \texttt{SBE} fields in the \texttt{mstatus} and \texttt{mstatush} registers as controlling the endianness of memory accesses, with the exception of instruction fetches, which are inherently little-endian. These fields have the following specific roles. \emph{Machine Endianness (\texttt{MBE})} governs the endianness for memory accesses in M-mode. Setting \texttt{MBE} to 0 enforces little-endian memory accesses while setting it to 1 enforces big-endian accesses. \emph{Supervisor Endianness (\texttt{SBE})} determines the endianness of memory accesses in S-mode, provided that S-mode is supported in the implementation. The \texttt{SBE} field is of particular importance for supervisor- and Hypervisor-level operations, such as page table management. 
However, a deviation from the expected behavior has been observed on the \texttt{RV64 CVA6} core. Specifically, manipulating the \texttt{MBE} or \texttt{SBE} fields in the \texttt{mstatus} register does not alter the endianness of explicit memory access instructions as anticipated. For instance, despite clearing or setting the \texttt{MBE} or \texttt{SBE} bits, the endianness of subsequent \texttt{sw} (store word) and \texttt{lb} (load byte) instructions remains unaffected. The issue can be reproduced via Listing~\ref{lst:v1_v2}.

\vspace{2mm}
\begin{lstlisting}[caption={Sample Code Snippet Demonstrating the Triggering of Vulnerabilities 1 and 2.}, label={lst:v1_v2}, numbersep=5pt,xleftmargin=2pt]
li t2, 0x12345678
sw t2, 0(t1)
lb t3, 0(t1)     // Exp(0x78) != Obs(0x78)
li t0, (1 << 37) // Set MBE (bit 37)
csrs mstatus, t0
sw t2, 0(t1)
lb t3, 0(t1)     // Exp(0x12) != Obs(0x78)
\end{lstlisting}

In the Listing ~\ref{lst:v1_v2}, we present code demonstrating the first vulnerability associated with \texttt{MBE}. Notably, the same code can be adapted to exploit a second vulnerability by altering the execution mode from M-mode to S-mode. This transition can be achieved by appropriately configuring the \texttt{mstatus} register followed by executing the \texttt{MRET} instruction. Additionally, the offset for setting the bit (\texttt{t0}) should be adjusted to $36$, corresponding to \texttt{SBE}, instead of $37$, which pertains to \texttt{MBE}.
However, in both vulnerability scenarios, the expected value from the last load instruction, Line 8, should be \texttt{0x12}, representing a big-endian load for \texttt{0x12345678}. Nonetheless, in both cases, the processor loads \texttt{0x78}, indicating that the endianness configuration did not change despite setting the \texttt{MBE} and \texttt{SBE} bits.

The vulnerabilities \texttt{V1} and \texttt{V2} are classified as severe due to their high CVSS 3.0 scores of $7.5$. These vulnerabilities present multiple attack vectors that adversaries can exploit to compromise the confidentiality, integrity, and availability of the system. Specifically, the processor fails to enforce the expected endianness for certain memory access instructions, such as \texttt{sw} and \texttt{lb}. An attacker can exploit this flaw by creating scenarios where discrepancies between the expected and actual behavior affect critical memory management structures, such as page tables. This exploitation can facilitate bypassing memory isolation mechanisms, leading to the corruption of page tables or unauthorized access to memory regions. Consequently, the attacker may gain elevated privileges or extract sensitive information. Furthermore, by combining these vulnerabilities with a software memory corruption attack, an attacker can destabilize the operating system kernel or hypervisor, resulting in arbitrary code execution or denial-of-service conditions.

\noindent\textbf{\texttt{V{3}} Improper Masking of Delegated Supervisor Timer Interrupts (STI) in CVA6 Processors.} 
According to the RISC-V ISA specification~\cite{riscv_spec_1,riscv_spec_2}, delegated interrupts should be masked at the delegator privilege level. Specifically, if the Supervisor Timer Interrupt (STI) is delegated to S-mode by setting the appropriate bit (5th bit in \texttt{mideleg}), the interrupt should not be taken when executing in M-mode. In this configuration, STIs should only trigger in S-mode, with control transferring to the corresponding S-mode interrupt handler. Conversely, if \texttt{mideleg[5]} is cleared, the interrupt is not delegated and should be taken in any mode, transferring control to the M-mode handler.

However, a deviation from the specified behavior has been observed in the \texttt{RV64 CVA6}~\cite{cva6} core. Specifically, when the STI is delegated to the S-mode by setting \texttt{mideleg[5]}, the interrupt remains visible in the M-mode, contrary to expectations. Even though the STI is delegated, it is still reflected in the \texttt{mip} register while executing in the M-mode rather than being masked. Instead, the interrupt should only appear in the \texttt{sip} register when executing in S-mode, indicating proper delegation. This behavior violates the RISC-V specification, as delegated interrupts should not appear at the delegator privilege level (in this case, M-mode). The issue can be reproduced using the PoC code snippet in Listing~\ref{lst:v3}.

\vspace{2mm}
\begin{lstlisting}[caption={Sample Code Demonstrating the Masking of Delegated STI.}, label={lst:v3}, numbersep=5pt,xleftmargin=2pt]
li t0, (1 << 5)// Load STI interrupt
csrs mip, t0   // Set STI in mip register
csrr t0, mip   // Check mask
csrr t0, sip   // Check delegation
\end{lstlisting}

The vulnerability \texttt{V3} is classified as severe due to its high CVSS 3.0 scores of $7.6$. Specifically, when the STI is delegated to the S-mode, it remains visible in the M-mode, violating the expected isolation between privilege levels defined by the RISC-V ISA. An attacker operating in S-mode can exploit this flaw by triggering repetitive STI interrupts and observing side effects in M-mode, such as execution timing variations or unexpected interrupt handling. This behavior allows the attacker to infer sensitive information about the M-mode firmware’s internal state, including interrupt handling logic, memory management operations, or context-switching activities.
Moreover, if M-mode software inadvertently processes these unexpected interrupts, it may expose unintended information or introduce vulnerabilities. For instance, M-mode may log the interrupts, update counters, or modify critical state variables, enabling an S-mode attacker to deduce whether M-mode is engaged in specific privileged operations or to infer details about memory layout and interrupt delegation mechanisms. Such leakage can be leveraged to bypass privilege isolation, potentially escalating the attacker’s privileges or providing a foothold for further exploitation.

\noindent\textbf{\texttt{V4} Improper Handling of \texttt{stval} Register in CVA6 Processors for \texttt{HFENCE.GVMA} Instruction.}  
According to the RISC-V ISA specification\cite{riscv_spec_1,riscv_spec_2}, when an illegal instruction exception occurs in Hypervisor/Supervisor Mode (HS-mode), the \texttt{stval} register can optionally return the faulting instruction bits. Specifically, if \texttt{stval} is written with a nonzero value when an illegal instruction exception occurs, it should contain the shortest of: a) the actual faulting instruction, b) the first ILEN bits of the faulting instruction, and c) the first SXLEN bits of the faulting instruction.

\vspace{2mm}
\begin{lstlisting}[caption={Sample Code Demonstrating the Improper stval Handling for HFENCE.GVMA}, label={lst:v4}, numbersep=5pt,xleftmargin=2pt]
li t0, (1 << 20)       // Load TVM
csrs mstatus, t0       // set TVM
/// Switch to HS-mode
hfence.gvma zero, zero // Exception here
\end{lstlisting}

Furthermore, in \texttt{HS-mode}, when the \texttt{mstatus.TVM} flag is set, executing the \texttt{HFENCE.GVMA} instruction should trigger an illegal instruction exception. Upon triggering the exception, the \texttt{stval} register should be set to the value of the faulting instruction. In this case, \texttt{stval} should hold the value of the \texttt{HFENCE.GVMA} instruction itself. However, a deviation from this expected behavior has been observed in the \texttt{CVA6 core}\cite{cva6}. Specifically, instead of setting \texttt{stval} to the correct faulting instruction value (\texttt{0x62000073}), the core erroneously sets it to \texttt{0x1}. This behavior violates the RISC-V specification, as the \texttt{stval} register is expected to contain the faulting instruction. The issue can be reproduced using the PoC code snippet in Listing~\ref{lst:v4}.
The vulnerability \texttt{V4} is classified as moderate (CVSS 3.0 score $5.5$), due to its potential impact on exception handling and debugging. This vulnerability could lead to uncertainty during debugging or exception handling. This, in turn, may result in incorrect exception handling behavior, particularly in complex systems that rely on accurate faulting instruction information for diagnostics or recovery procedures. In systems where accurate fault reporting is crucial for security or stability, this could introduce difficulties in identifying the root cause of the exception, potentially affecting system reliability or security.

\noindent\textbf{\texttt{V5} Access Control Issue in CSR Register Files.}
During the discussion with the CVA6 developers regarding vulnerabilities \texttt{V1} and \texttt{V2}, we realized a newer vulnerability that can be interpreted by \texttt{V1} and \texttt{V2}, which could be far more critical. This issue represents a critical access control problem in the Control Status Register (CSR) register files module. Specifically, the \texttt{MBE} and \texttt{SBE} bits, which are expected to be read-only and set to zero, can be modified, contrary to the CVA6 specification\cite{cva6_endian}. As demonstrated in Listing~\ref{lst:v1_v2}, the \texttt{MBE} and \texttt{SBE} bits could be changed, despite being expected to be locked at zero.
The vulnerability \texttt{V5} is classified as severe (CVSS 3.0 score $7.6$),
due to its potential impact on CSR access control, as it allows for unintended modification of critical status information that is supposed to be protected.

\nadd{\noindent\textbf{Two New Vulnerabilities on a Commercial Core.} \ourname successfully detected two bugs in the implementation of extensions of Beyond's BA51-H core~\cite{pinto2024novel}, a commercialized design developed during the CROSSCON project~\cite{crosscon}. For confidentiality reasons, we cannot disclose the full bug details. However, one of the identified issues appeared while accessing specific registers, highlighting a subtle interaction flaw within the newly developed functionalities. This discovery demonstrates \ourname's capability to uncover critical design flaws in evolving commercial hardware, showcasing its significant value for industrial applications.
}

\section{Ablation and Hyperparameter Study}
\label{sec:Ablation Study}
\begin{figure*}[htbp]
    \centering
     \subfloat[The Need for the GRM Exploration.]{%
        \includegraphics[scale=0.33]{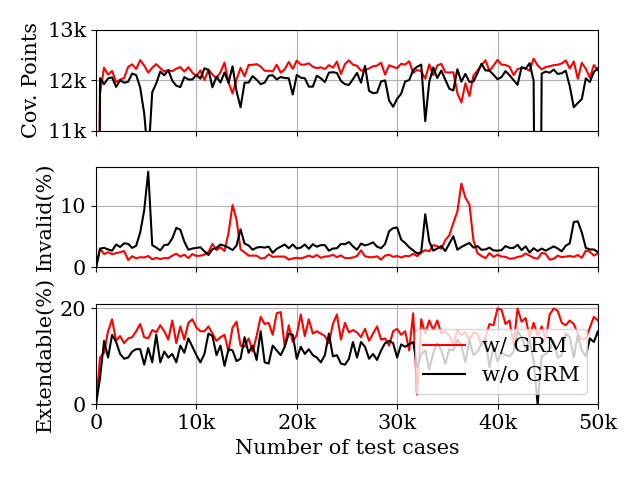}
        \label{fig:ablation_pretraining}}
     \subfloat[Instruction Block Settings.]{
        \includegraphics[scale=0.33]{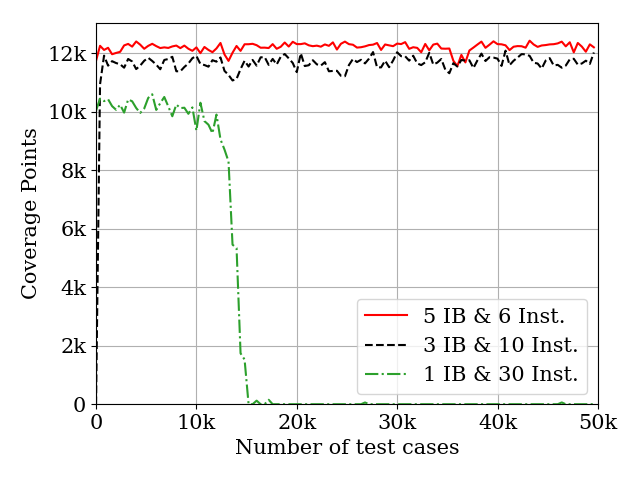}
        \label{fig:ablation_block_inst}}
      \subfloat[Robustness of \ourname.]{%
        \includegraphics[scale=0.33]{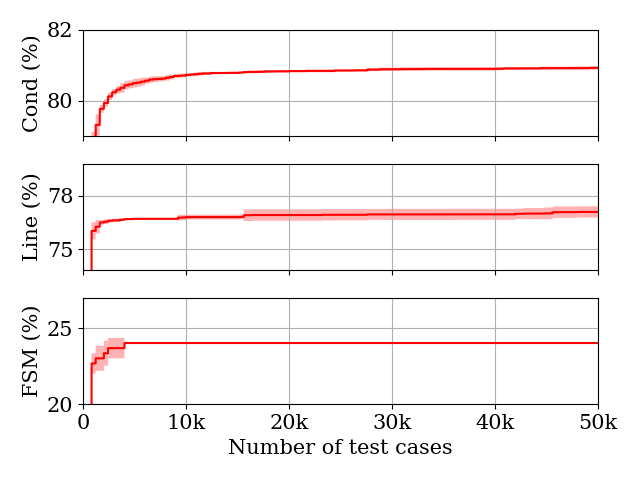}
        \label{fig:ablation_robustness}}
    \caption{Ablation and Hyperparameter Studies on Critical Settings.}\label{fig:ablation_study}
\end{figure*}

\subsection{The Need for GRM Fuzzing}
\nadd{As the core component of \ourname, GRM plays a crucial role in reducing overhead within the GoldenFuzz framework by acting as a ``digital twin'' of the DUT.} To assess the effectiveness and necessity of this component, we conduct an ablation study on the performance of \ourname with and without GRM.

Figure~\ref{fig:ablation_pretraining} presents the results across three key metrics: 1) coverage points, which aggregate condition, line, and finite-state machine (FSM) coverage; 2) invalid rate, which measures the proportion of IBs that are syntactically invalid; and 3) extendable rate, which quantifies the percentage of IBs that are suitable for concatenation in future iterations. As defined in Section~\ref{subsec:pipeline}, an IB is considered extendable (or, not ``dead'') if it is syntactically valid and fully executable, or largely so in the case of branch instructions. Across all three metrics, GRM-guided fuzzing (red curve) consistently outperforms the baseline that explores solely the DUT. Specifically, GRM fuzzing achieves higher overall coverage, generates fewer invalid IBs, and yields a greater proportion of extendable IBs. These results highlight a critical advantage of using a GRM during early-stage policy refinement: it enables safer and more efficient coverage exploration without risking the stability or limited observability of the DUT. \nadd{From a higher level, by leveraging the GRM for initial, rapid, and low-cost coverage-guided test case refinement before targeting the actual hardware, GoldenFuzz significantly accelerates deep architectural exploration and vulnerability discovery.}

\subsection{The Selection of Instruction Blocks Settings.}
\label{subsec:The Selection of Instruction Blocks Settings}
\nadd{Instruction block-based testcase generation enable \ourname to steadily explore new hardware states with low computation overhead.}
To evaluate the impact of instruction block (IB) granularity on fuzzing effectiveness, we compare three block configurations: five IBs of six instructions each (5-6), three IBs of ten instructions (3-10), and one IB containing all 30 instructions (1-30). 
The y-axis in Fig.~\ref{fig:ablation_block_inst} shows the hardware coverage points achieved over learning iterations.

The results demonstrate that using more, smaller IBs (5-6) consistently leads to higher coverage. This improvement stems from two key advantages. First, smaller blocks offer finer-grained exploration, allowing each IB to independently probe different hardware states. This increases the likelihood of reaching diverse execution paths and reduces the risk of coverage stagnation. Second, block-wise generation significantly eases the learning task for the fuzzer. Indeed, the fuzzer only needs to learn how to construct shorter, more manageable IBs, reducing the search space and accelerating convergence.

In contrast, larger IBs (e.g., 1-30) are more prone to becoming dead, i.e., failing to execute due to syntax or runtime errors such as invalid instructions or premature termination. Indeed, longer IBs have a higher chance of being discarded, which reduces the pool of viable test cases and leads to lower overall coverage. This degradation can reach a point where the system fails to generate any valid IBs at all. 

While configurations like 30-1 (30 one-instruction IBs) may offer greater stability due to minimal instruction complexity per block, they introduce significant computational overhead and reduce the contextual expressiveness of each block. Our experiments show that a middle ground, such as the 5-6 configuration, offers the best trade-off between learning efficiency, coverage growth, and computational feasibility.

\subsection{Robustness of \ourname.}
\label{subsec:robustness}
To evaluate the robustness and consistency of our fuzzing strategy, we conducted five independent fuzzing runs and measured the resulting coverage across three metrics: line, coverage, and FSM coverage. Figure~\ref{fig:ablation_robustness} presents the mean coverage and its variance (shaded regions) as the number of test cases increases; the y-axis scales are set differently for better visibility. Across all three metrics, \ourname demonstrates stable and consistent behavior. Indeed, the narrow variance bands across all metrics demonstrate that \ourname is not only effective but also robust, consistently achieving high architectural coverage regardless of initialization or randomness in the fuzzing process.
\section{Discussion}
\label{sec:discussion}

\noindent
\textbf{Cross-design Generalization.}
A key advantage of \ourname is its portability across implementations of the same ISA. In our evaluation across three distinct RISC-V cores, including the Out-of-Order BOOM processor~\cite{boom}, no changes were required to the framework. 
Indeed, all ISA-compliant processors, regardless of their microarchitecture, must expose the same architectural state at instruction retirement. Consequently, \ourname is directly applicable to any conforming RISC-V core. 
\nadd{Adapting \ourname to other ISAs (e.g., ARM or x86) and more complex (closed-source) designs follows the same workflow. 
While technically feasible, the practical limitations are primarily tied to (1) the availability of GRM and (2) the closed-source nature of ISA and design implementation. Recall that \ourname relies on GRM for the low-cost fuzzing policy refinement and, like other white-box fuzzers~\cite{rostami2024beyond,solt2024cascade,kande2022thehuzz,hur2021difuzzrtl}, leverages RTL-level coverage feedback to explore deeper hardware states. The lack of GRM would degrade \ourname to a conventional fuzzer that directly interacts with DUT; the lack of design insight further degrades \ourname into a pure random fuzzer and poses a significant challenge to vulnerability detection. These availability and open-source constraints reflect a broader challenge in applying advanced fuzzing techniques to commercial and closed-source platforms.}

\noindent
\nadd{\textbf{LLM Advances for \ourname.}
Recent advances in Large Language Models (LLMs) offer promising avenues for further improving hardware vulnerability detection. Incorporating techniques like Retrieval-Augmented Generation (RAG)~\cite{lewis2020retrieval} could enhance hardware fuzzing by integrating design specifications, thereby increasing the efficiency of vulnerability detection. For instance, an LLM could retrieve relevant design documentation to inform its test case generation, leading to more targeted and effective fuzzing. Additionally, allowing the fuzzer to understand human languages opens the opportunities of applying prompt engineering techniques to generate more refined test cases specifically tailored for, e.g., targeted building blocks within the hardware design or certain CWEs. These advancements could lead to more intelligent and context-aware fuzzing strategies, thus uncovering critical vulnerabilities.}

\section{Related Work}
\label{sec:related}

Hardware fuzzing frameworks have been widely deployed to various designs, such as SoCs, CPUs, and isolated IP blocks. We categorize them into generic fuzzers and processor fuzzers.

\noindent
\textbf{Generic Fuzzers.} Existing works, such as Laufer et al.~\cite{laeufer2018rfuzz} and Li et al.~\cite{li2021symbolic}, demonstrate the feasibility of using FPGA emulation-based generic fuzzing based on multiplexer control signals. However, these approaches are reliant on specific hardware design languages (e.g., HDL), limiting their scalability. Further, the overheads in monitoring multiplexers in complex designs hamper the usability~\cite{canakci2023processorfuzz}. 
In contrast, Trippel et al.~\cite{trippel2022fuzzing} proposed fuzzing hardware-like software by fuzzing the hardware simulation binary rather than porting software fuzzers directly on the hardware designs. Verilator~\cite{Verilator} translates the hardware to equivalent software. This approach allows fuzzing to utilize existing software coverage metrics, such as basic block and edge coverage~\cite{canakci2023processorfuzz}. Still, it faces scalability problems when, e.g., fuzzing a whole CPU.

\noindent
\textbf{Processor Fuzzers.} 
TheHuzz~\cite{kande2022thehuzz} simulates the RTL design of the processor with the binary format of the instruction using Synopsys VCS~\cite{vcs} that traces code coverage through various metrics, including branch, condition, toggle, FSM, and functional coverage. However, this method suffers from low computation efficiency and hardware coverage.
DifuzzRTL~\cite{hur2021difuzzrtl} generates instructions and collects control register coverage to guide the fuzzing process. Following this work, MorFuzz~\cite{xu2023morfuzz} achieves a final coverage that is 4.4 times higher than DifuzzRTL by generating fuzzing seeds based on syntax and semantics and using runtime information feedback to mutate instructions. 
ProcessorFuzz~\cite{canakci2023processorfuzz} is a concurrent work that generates instructions and collects coverage of control and status registers. However, these works only focus on the coverage of registers generating the select signals of MUXes, leading to missing bugs and vulnerabilities.
To increase the design coverage and detect more vulnerabilities, HyPFuzz~\cite{chen2023hypfuzz} was proposed to guide the fuzzer by the formal verification tools reaching hard-to-reach design spaces. Alternatively, SoC Fuzzer~\cite{hossain2023socfuzzer} directs the fuzzing based on the security properties (generic cost function) that detect vulnerabilities in the DUT. Finally, Cascade~\cite{solt2024cascade} aims to enhance instruction execution efficiency by constructing long programs and eliminating control flow influences. It conducts the entire fuzzing process at the program-level granularity without using any mutation strategies to guide fuzzing. However, these approaches overlook the complexity of input semantics and receive limited feedback from the DUT. Unfortunately, overly long and complex test cases lead to high time consumption; the basic block's usage constrains the test case's variety, thus reducing the bug detection capability.
ChatFuzz~\cite{rostami2024beyond} uses a pre-trained language model, fine-tuned by reinforcement learning on instruction binaries, for the hardware fuzzing. In contrast, we specifically customize a GPT model with the proposed hardware fuzzing scheme, leading to lighter computation overhead and significantly better performance in hardware coverage and vulnerability detection.

\section{Conclusion}
\label{sec:conclusions}
In this paper, we present \ourname, a novel language-model-based hardware fuzzer that decouples test case refinement from coverage and vulnerability exploration. It first uses a software-based Golden Reference Model (GRM), a “digital twin” conforming to the DUT’s ISA, to efficiently refine fuzzing strategies before executing targeted tests on the actual DUT, reducing the cost for cycle-accurate simulations.
Empirical results show that \ourname outperforms state-of-the-art fuzzers in both coverage and computational cost. It detects all previously known vulnerabilities and uncovers seven new ones in open-source and commercial cores, four of which are rated highly severe (CVSS $>$ 7.0), highlighting their real-world impact.

\section*{Acknowledgement}
\label{sec:ccknowledgement}
Our research work was partially funded by Intel’s Scalable Assurance Program, DFG-SFB 1119-236615297, the European Union under Horizon Europe Programme-Grant Agreement 101070537-CrossCon, NSF-DFG-Grant 538883423, the European Research Council under the ERC Programme-Grant 101055025-HYDRANOS, and Synopsys (special thanks to Catherine Le Lan) with EDA tools licenses support. This work does not in any way constitute an Intel endorsement of a product or supplier. Any opinions, findings, conclusions, or recommendations expressed herein are those of the authors and do not necessarily reflect those of Intel, the European Union, or the European Research Council.

\newpage
\section*{Ethics Considerations}
\ourname is a hardware fuzzing tool developed to advance the functional and security verification of hardware, particularly processors. Its intended users include security researchers, hardware manufacturers and designers, and hardware security companies. By using the \ourname framework, users can discover new bugs and vulnerabilities in hardware designs under test. We thoroughly evaluated \ourname on three widely recognized benchmarks, discovering \newbugcount{} new bugs and vulnerabilities. In line with the Menlo Report principles\cite{menlo_report}, particularly the principle of "Respect for Persons", we have ensured that the security vulnerabilities identified by \ourname were promptly communicated to the responsible teams for CVA6~\cite{cva6}, BOOM~\cite{boom}, and RocketChip~\cite{rc}. This timely disclosure was essential to mitigate risks and protect individuals from potential harm that could arise if adversaries were to uncover these vulnerabilities independently. The responsible parties have acknowledged the issues and are actively working on implementing fixes to address the concerns raised in this paper.

To further adhere to the principle of justice, which requires fairness in distributing benefits and burdens, we have carefully considered the potential risks associated with the misuse of \ourname. To prevent harm and ensure that the framework is used to advance research and enhance hardware security, access to the source code for \ourname will be restricted. It will be made available only upon request and with confirmation that it will be used exclusively for responsible research by academic users and hardware manufacturers. This approach aligns with the Menlo Report~\cite{menlo_report} emphasis on promoting social value while protecting individual rights and privacy, ensuring that \ourname contributes positively to the hardware security community without introducing risks.


\bibliographystyle{IEEEtran}
\bibliography{bibliography}

\end{document}